\documentstyle[11pt,epsfig]{article}
\begin{document}
\newcommand{\roughly}[1]%


\newcommand{\PSbox}[3]{\mbox{\rule{0in}{#3}

\includegraphics{#1}\hspace{#2}}}
\newcommand\lsim{\roughly{<}}
\newcommand\gsim{\roughly{>}}
\newcommand\CL{{\cal L}}
\newcommand\CO{{\cal O}}
\newcommand\half{\frac{1}{2}}
\newcommand\beq{\begin{eqnarray}}
\newcommand\eeq{\end{eqnarray}}
\newcommand\eqn[1]{\label{eq:#1}}
\newcommand\intg{\int\,\sqrt{-g}\,}
\newcommand\eq[1]{eq. (\ref{eq:#1})}
\newcommand\meN[1]{\langle N \vert #1 \vert N \rangle}
\newcommand\meNi[1]{\langle N_i \vert #1 \vert N_i \rangle}
\newcommand\mep[1]{\langle p \vert #1 \vert p \rangle}
\newcommand\men[1]{\langle n \vert #1 \vert n \rangle}
\newcommand\mea[1]{\langle A \vert #1 \vert A \rangle}
\newcommand\bi{\begin{itemize}}
\newcommand\ei{\end{itemize}}
\newcommand\be{\begin{equation}}
\newcommand\ee{\end{equation}}
\newcommand\bea{\begin{eqnarray}}
\newcommand\eea{\end{eqnarray}}

\def\Dsl{\,\raise.15ex \hbox{/}\mkern-12.8mu D}
\newcommand\Tr{{\rm Tr\,}}
\thispagestyle{empty}

\begin{titlepage}



\vspace{1.0cm}

\begin{center}

{\LARGE \bf  Correlated Random Walks and the Joint Survival Probability}\\

\bigskip

\bigskip

\bigskip

{Mark B. Wise$^a$ and Vineer Bhansali$^b$} \\

~\\

\noindent

{\it\ignorespaces

          (a) California Institute of Technology, Pasadena CA 91125\\

           {\tt wise@theory.caltech.edu}\\

\bigskip   
(b) PIMCO, 840 Newport Center Drive, Suite 300\\

               Newport Beach, CA 92660 \\

{\tt   bhansali@pimco.com}

}\bigskip

\end{center}

\vspace{1cm}

\begin{abstract}
First passage models, where corporate assets undergo correlated random walks
and a company defaults if its assets fall below a threshold provide
an attractive framework for modeling the default process. Typical one year default correlations are small, i.e., of order a few percent, but nonetheless including correlations is very important, for managing portfolio credit risk and pricing some credit derivatives (e.g. first to default baskets). In first passage models the exact dependence of the joint survival probability of more than two firms on their asset correlations is not known. We derive an expression for the dependence of the joint survival probability of $n$ firms on their asset correlations using first order perturbation theory in the correlations. It includes all terms that are linear in the correlations but neglects effects of quadratic and higher order. For constant time independent correlations we compare the first passage model expression for the joint survival probability with what a multivariate normal Copula function gives.  As a practical application of our results we calculate the dependence of the five year joint survival probability for five basic industrials on their asset correlations.
\end{abstract}

\vfill


\end{titlepage}


\section{Introduction}

In the management of credit risk and the pricing of corporate bonds and credit derivatives a fundamental role is played by  corporate default and survival probabilities. Structural first passage models [1-7] are one of the standard approaches to estimating these probabilities. In such models the firm asset value evolves with time in a random walk and default occurs if the value of the firm's assets fall below a default threshold. Merton [1] showed that corporate stock and debt are options on the firm asset value. Equity holders hold a call option on the firm asset value, while bond holders are short a put on the firm asset value. In first passage models a simple analytic expression can be derived for the risk neutral probability of a company surviving without defaulting during a time interval $t$, $P(t)$ \footnote{ The risk neutral probability of the company defaulting during the time  $t$ is then $1-P(t)$.}. Risk neutral probabilities are used for pricing corporate bonds and credit derivatives.  A no arbitrage argument implies they are calculated by taking the drift in the firm value to be given by the risk free rate of return rather than its ''real world'' value.

For portfolios of bonds the joint survival probabilities are needed for a proper assessment portfolio risk. In addition the pricing of some credit derivatives (e.g., first to default baskets) depends on the risk neutral joint survival probabilities. These can be calculated using a first passage model with correlated random walks or by postulating a particular Copula function.  A multivariate normal Copula function is often used. In first passage models the exact dependence of the joint survival probability of more than two firms on their asset correlations is not known. We derive a formula for the joint survival probability that includes the effects of asset correlations treating the correlations to linear order in perturbation theory. For constant time independent correlations we compare this expression for joint survival probabilities with what a multivariate normal Copula function gives.  At linear order in perturbation theory in the asset correlations the joint survival probability of $n$ firms can be expressed in terms of the firms default correlations.

The perturbation theory results derived in this paper for the joint survival probability are valid provided the asset correlations and the number of firms are not too large. We argue that for five firms perturbation theory provides a useful approximation to the joint survival probability for realistic values of the asset correlations. When the number of firms is very large our results provide a limit (i.e., the small correlation limit) where numerical simulations of correlated random walks, with first passage boundary conditions, can be compared with analytic results.

As a practical application of our results we calculate the dependence of the five year risk neutral joint survival probability of five basic industrials on their asset correlations. The companies we consider are: Alcoa Inc., Dow Chemical Company, E.I. du Pont de Nemours and Company, International Paper Company, and Weyerhauser Company. 

\section{Joint Survival Probabilities}

The asset value $V_i$ of a company $i$ is the sum of the value of its stock $S_i(V_i,t)$ and its debt $B_i(V_i,t)$. Assuming dividend payments to stock holders of company $i$, $q_iV_i{\rm d}t$ in a time interval ${\rm d}t$, we take the firm asset values of the $n$-firms (i.e., $i=1,\ldots,n)$ to evolve with time according to 
\be
\label{assetevolv}
{\rm d}V_i=\mu_i V_i {\rm d}t +V_i\sum_{j=1}^n\sigma_{ij}{\rm d}w_j- q_i V_i {\rm d}t,
\ee
where $w_i$ are standard Brownian motions, i.e.,
\be
\label{brown}
 E[{\rm d}w_i {\rm d}w_j]={\rm d}t\delta_{ij}
\ee
and the Kronecker delta is defined by $\delta_{ij}=1$ if $i=j$ and $\delta_{ij}=0$ if $i \ne j$. In Eq.(\ref{assetevolv}) $\mu_i$ is the drift of the $i$'th company value if there are no dividend payments to its stock holders\footnote{To simplify the discussion in this section, we take the debt to be zero coupon bonds that mature at a date that is beyond our investment time horizon.}. Here we are assuming that dividend payments are financed out of operating profits or cash reserves and not out of the issuance of new debt or stock. We also assume that the dividend yields $q_i$ are constant independent of time. Time depended dividend payments were studied in [8]. The  assumptions made in this section are consistent with the Modigliani-Miller theorem [9,10].

Eqs.~(\ref{assetevolv}) and (\ref{brown}) imply that the asset (return) covariance matrix $\rho_{ij}$ is
\begin{equation}
\rho_{ij}=\sum_{k=1}^n \sigma_{ik}\sigma_{jk}.
\end{equation}
Hence the asset volatilities are
\be
\sigma_i=\sqrt{\sum_{k=1}^n \sigma_{ik}\sigma_{ik}},
\ee
and the asset correlation matrix is
\be
\xi_{ij}={1 \over \sigma_i \sigma_j}\sum_{k=1}^n \sigma_{ik}\sigma_{ik}.
\ee
We allow the default barrier to have an exponential time dependence and assume that company $i$ defaults at a time $t$ if its firm value falls below,
\be
\label{defthres}
V_{id}=d_i~e^{\lambda_i t},
 \ee
where $d_i$ and $\lambda_i$ are constants independent of time. This is the framework of Black and Cox [2]. One usually takes the exponential dependence to be given by the risk free rate $r$, i.e., $\lambda_i=r$.  This form for $V_{id}$ could arise from a covenant that firm $i$ has with its debt holders which forces the firm into bankruptcy to ensure that the debt holders do not lose more than a fixed fraction of the present value of their principal. 

The default probabilities follow from the expressions for the evolution of the firm value and the default barrier given in Eqs.~(\ref{assetevolv}) and (\ref{defthres}). The constant real world asset drift $\mu_i$ in Eq.~(\ref{assetevolv}) does not affect $S_i(V_i,t)$ and $B_i(V_i,t)$. Risk neutral default and survival probabilities that are used for pricing corporate bonds and credit default swaps are obtained by setting the asset drift $\mu_i$ equal to the risk free rate $r$.

It is convenient to change variables from the firm value $V_i$ to $z_i$ defined by,
\be
\label{newvar} 
z_i(V_i,t)=({\rm log}(V_i/V_{i0}) - \lambda_i t).
\ee
Default occurs if $z_i$ falls below the time independent default threshold
 \be
z_{id}={\rm log}(d_i/V_{i0}).
 \ee
In Eq.~(\ref{newvar}) $V_{i0}$ is the initial asset value for firm $i$. Ito's lemma implies that the time
evolution of $z_i$ is given by,
 \be \label{relation}
{\rm d}z_i=\eta_i {\rm d}t+\sum_{j=1}^n \sigma_{ij}  {\rm d}w_j,
\ee 
where
\be
\eta_i=\mu_i-{1 \over 2}\sigma_i^2-\lambda_i-q_i
\ee
$V_i$ can be expressed in terms of $z_i$ and $t$ using
Eq.~(\ref{newvar}).

Let $p(z_1,\ldots, z_n,t) {\rm d}z_1\cdots {\rm d}z_n$ be the joint probability that the $n$ companies survive to time $t$ (i.e., none of them default before time $t$) with each $z_i$ in the interval ${\rm d}z_i$. More explicitly,
\be
p(z_1,\ldots, z_n,t) {\rm d}z_1\cdots {\rm d}z_n=P[t_{\rm default}>t ~~{\rm for~ all}~n~{\rm firms} |z_i(t) \in {\rm d}z_i].
\ee
The probability that all n companies survive to time t (i.e., the joint survival probability) is given by
\be
P_{1, \ldots, n}(t)=\int_{z_{1d}}^{\infty} {\rm d}z_1\cdots \int_{z_{nd}}^{\infty}{\rm d}z_n p(z_1,\ldots, z_n,t).
\ee

Since $z_i$ evolves via Eq~(\ref{relation}) the probability density $p(z_1,\ldots, z_n,t)$
obeys the diffusion differential equation (see, for example, [11]),
\be
\label{partia}
\left({\partial  \over \partial t}+\sum_i\eta_i {\partial  \over \partial z_i}-{1 \over 2}\sum_{i,j}\rho_{ij}{\partial^2  \over \partial z_i \partial z_j}\right) p(z_1,\ldots, z_n,t)=0.
\ee
where the sums go over all $n$ values of $i$ and $j$. The probability density $p(z_1,\ldots, z_n,t)$ obeys the initial condition\footnote{$\delta(x)$ denotes the Dirac delta function. It is defined by, $\delta (x)=0$ for $x\ne 0$ and $\int \delta(x) f(x)=f(0)$ for any smooth function $f(x)$.} (here $z_i > z_{id}$)
\be
p(z_1,\ldots, z_n,0)=\delta(z_1)\cdots \delta(z_n),
\ee
and the first passage boundary conditions,
\begin{eqnarray}
p(z_{1d},z_2, \ldots z_n,t)=\ldots=p(z_1, \ldots,z_{n-1}, z_{nd},t)&=& 0\nonumber \\
p(\infty,z_2, \ldots z_n,t)=\ldots=p(z_1, \ldots,z_{n-1}, \infty,t)&=&0.
\end{eqnarray}

It is convenient to reexpress this in terms of the volatilities and the correlation matrix,
\be
\label{parti}
\left({\partial  \over \partial t}+\sum_i\eta_i {\partial  \over \partial z_i}-{1 \over 2}\sum_i \sigma_i^2{\partial^2  \over \partial z_i^2}-{1 \over 2} \sum_{i \ne j}\sigma_i \sigma_j \xi_{ij}{\partial^2  \over \partial z_i \partial z_j}\right)p(z_1,\ldots, z_n,t)=0.
\ee
For $n=2$ the solution to this differential equation with the appropriate boundary conditions for the first passage problem has been expressed in terms of an infinite series [12]. Here we consider the case where the off diagonal elements of the correlation matrix are small enough to treat them as a perturbation. Expanding the joint survival probability density in powers of the off diagonal correlation matrix elements we write
\be
p(z_1,\ldots, z_n,t)=p^{(0)}(z_1,\ldots, z_n,t)+p^{(1)}(z_1,\ldots, z_n,t)+ \ldots,
\ee
where $p^{(0)}(z_1,\ldots, z_n,t)$ is the solution with the asset correlations set to zero, $p^{(1)}(z_1,\ldots, z_n,t)$ contains all the terms linear in the off diagonal elements of the correlation matrix, etc. A similar expansion is made for the joint survival probability itself
\be
P_{1, \ldots, n}(t)=P^{(0)}_{1,\ldots, n}(t)+P^{(1)}_{1, \ldots , n}(t)+\ldots .
\ee
We consider the case where the volatilities and correlations are constants independent of time. Then
\be
p^{(0)}(z_1,\ldots, z_n,t)=p_1(z_1,t) \cdots p_n(z_n,t),
\ee
where the probability density $p_j(z_j,t)$ is defined so that $p_j(z_j,t)dz_j$ is the probability that the company $j$ survives to time $t$ (i.e., does not default before time $t$) with $z_j$ in the interval ${\rm d}z_j$. It is given by
\be
p(z_j,t)={1 \over {\sqrt{2 \pi \sigma_j^2 t}}}\left[ e^{-(z_j-\eta_j t)^2/(2\sigma_j^2 t)}-e^{2 \eta_j z_{jd}/\sigma_j^2} e^{-(z_j-2z_{jd}-\eta_j t)^2/(2\sigma_j^2 t)} \right].
\ee
The probability of company $j$ surviving to time t is the integral of this over allowed values of $z_j$ and it is given by
\be
\label{probint}
P_{j}(t)={1 \over 2}\left({ \rm erf}\left[{\eta_j t-z_{jd} \over \sqrt{2 \sigma_j^2 t}}\right]+1 \right)-{1 \over 2}e^{2 \eta_j z_{jd}/\sigma_j^2}\left({ \rm erf}\left[{\eta_j t+z_{jd} \over \sqrt{2 \sigma_j^2 t}}\right]+1 \right),
\ee
where the error function is defined by
 \be
 {\rm erf}(z) \equiv {2\over \sqrt{\pi}} \int_0^z {\rm d}x ~e^{-x^2}.
 \ee
The leading joint survival probability is
\be
P^{(0)}_{1, \ldots , n}(t)=P_{1}(t) \cdots P_{n}(t).
\ee

The contribution to the joint survival probability density that is linear in the asset correlations $p^{(1)}(z_1,\ldots, z_n)$ satisfies the differential equation
\be
\label{partib}
\left({\partial  \over \partial t}+\sum_i\eta_i {\partial  \over \partial z_i}-{1 \over 2}\sum_i \sigma_i^2{\partial^2  \over \partial z_i^2}\right)p^{(1)}(z_1,\ldots, z_n,t)={1 \over 2} \sum_{i \ne j}\sigma_i \sigma_j \xi_{ij}{\partial^2  \over \partial z_i \partial z_j}p^{(0)}(z_1,\ldots, z_n,t)=0.
\ee
The Greens function (see, for example, [13] for a review of perturbation theory techniques) for the diffusion equation, $K(z_1, \ldots ,z_n,x_1, \ldots , x_n;t-t')$, satisfies the differential equation
\be
\left({\partial  \over \partial t}+\sum_i\eta_i {\partial  \over \partial z_i}-{1 \over 2}\sum_i \sigma_i^2{\partial^2  \over \partial z_i^2}\right)K(z_1, \dots, z_n,x_1, \dots x_n,;t-t')=\delta(t-t')\delta(z_1-x_1)\cdots \delta(z_n-x_n).
\ee

The solution to the differential equation for $p^{(1)}(z_1,\ldots, z_n,t)$ is
\begin{eqnarray}
&&p^{(1)}(z_1,\ldots, z_n,t)= {1 \over 2} \sum_{i \ne j} \int_0^{\infty} {\rm d}t'  \sigma_i \sigma_j \xi_{ij}\nonumber \\
&\times&  \int_{x_{1d}}^{\infty}{\rm d}x_1 \cdots \int_{x_{nd}}^{\infty}{\rm d}x_n K(z_1, \dots, z_n,x_1, \dots x_n,;t-t'){\partial^2  \over \partial x_i \partial x_j}p^{(0)}(x_1,\ldots, x_n,t')
\end{eqnarray}
The Greens function satisfying first passage boundary conditions is
\be
\label{green1}
K(z_1, \dots, z_n,x_1, \dots x_n,;\tau)=\theta(\tau)p(z_1,x_1;\tau)\cdots p(z_n,x_n;\tau),
\ee
where $\theta(\tau)$ is defined to be unity for $\tau \ge 0$ and zero otherwise and 
\be
\label{green2}
p(z_j,x_j;\tau)={1 \over {\sqrt{2\pi \sigma_j^2\tau}}}\left[e^{-(z_j-x_j-\eta_j \tau)^2/(2 \sigma_j^2 \tau)}-e^{-(z_j+x_j-2z_{jd}-\eta_j \tau)^2/(2\sigma_j^2 \tau)+2\eta_j(z_{jd}-x_j)/\sigma_j^2} \right].
\ee
Note that $p(z_j,x_j;\tau){\rm d}z_j$ is the probability that the asset value of company $j$ starts off at $x_j$ and random walks without defaulting in the time $\tau$ to a region ${\rm d}z_j$ about the value $z_j$. Hence $p(z_j,0;\tau)=p(z_j,\tau)$.
Using Eqs.~(\ref{green1}) and (\ref{green2})
\begin{eqnarray}
&&p^{(1)}(z_1,\ldots, z_n,t)={1 \over 2}\sum_{i \ne j}\left(\prod_{k \ne i,j}p_k(z_k,t)\right) \int_0^t{\rm d}t'\sigma_i\sigma_j\xi_{ij}  \nonumber \\
&&\times\left(\int_{x_{id}}^\infty {\rm d}x_i p(z_i,x_i;t-t'){\partial \over \partial x_i}p(x_i,t')\right)\left(\int_{x_{jd}}^\infty {\rm d}x_j p(z_j,x_j;t-t'){\partial \over \partial x_j}p(x_j,t')\right)
\end{eqnarray}

We are interested in the effects of the asset correlations on the joint survival probability. Using the above results we find that
\be
\label{mainresult1}
P^{(1)}_{1, \ldots, n}(t)/P^{(0)}_{1, \ldots, n}(t)= {1 \over 2}\sum_{i \ne j}{1 \over P_{i}(t) P_{j}(t)} \int_0^t{\rm d}t' \sigma_i\sigma_j\xi_{ij} A_{ij}(t,t'),
\ee
where,
\be
\label{mainresult2}
A_{ij}(t,t')=\left(\int_{x_{id}}^\infty {\rm d}x_i U_i(x_i;t-t'){\partial \over \partial x_i} p_i(x_i,t')\right) \left(\int_{x_{jd}}^\infty {\rm d}x_j U_j(x_j;t-t'){\partial \over \partial x_j} p_i(x_j,t')\right),
\ee
and
\begin{equation}
\label{mainresult3}
U_j(z_j;\tau)={1 \over 2}\left({\rm erf}\left[{\eta_j \tau -z_{jd}+z_j \over \sqrt{2 \sigma_j^2\tau}}\right]+1\right)-{1 \over 2}{\rm e}^{2\eta_j(z_{jd}-z_j)/\sigma_j^2 }\left({\rm erf}\left[{\eta_j \tau +z_{jd}-z_j \over \sqrt{2 \sigma_j^2\tau}}\right]+1\right).
\end{equation}
Eqs.~(\ref{mainresult1}), (\ref{mainresult2}), and (\ref{mainresult3}) are the main results of this paper. They hold even in the case of time dependent correlations with the replacement $\xi_{ij} \rightarrow \xi_{ij}(t')$ in Eq.~(\ref{mainresult1}). It is possible to derive the expression for $P^{(2)}_{1, \ldots, n}$ (i.e, the correction to the joint survival probability at second order in the correlations) however the number of numerical integrations required to compute it increases (from three) to six.

Note that Eq.~(\ref{mainresult1}) implies that at linear order in the correlations the joint survival probability of $n$ firms is expressible as a linear combination of the joint survival probability of pairs of firms. Explicitly,
\be
\label{toobad}
P_{1,\ldots, n}(t)=P_1(t) \cdots P_n(t)\left[1+{1 \over 2} \sum_{i \ne j}\left({P_{i,j}(t) \over P_i(t)P_j(t)}-1 \right)\right].
\ee
Thus, for example, $P_{1,2,3}(t)=P_{1,2}(t)P_3(t)+P_{1,3}(t)P_2(t)+P_{2,3}(t)P_1(t)-2P_1(t)P_2(t)P_3(t)$.

It is sometimes convenient to introduce the random variables $\hat n_i(t)$ that take the value 1 if firm $i$ survives without defaulting until time $t$ and zero otherwise. Expectations of products of these variables are equal to the joint survival probabilities, $E[\hat n_1(t) \cdots \hat n_n(t)]=P_{1,\ldots , n}(t)$. Fluctuations about the expected value of these variables are characterized by,  $\delta \hat n_i(t)= \hat n_i(t)-E[\hat n_i(t)]$. Eq.~(\ref{toobad}) implies that 
\be
\label{skewvan}
E[\delta \hat n_i(t)\delta \hat n_j(t)\delta \hat n_k(t)]=0.~~~~~i \ne j \ne k .
\ee
To understand the implications of Eq.~({\ref{skewvan}) for portfolio risk consider a portfolio of corporate bonds subject to default risk and take the investment time horizon to be $ t$. For simplicity we assume an initial investment of one dollar in the bonds of firm $i$ has, after time $t$, the value $1+c_i$ dollars if firm $i$ doesn't default and value $R_i+c_i$ if firm $i$ does default. $R_i$ is the recovery fraction and $c_i$ is the promised corporate return. Then the random variable that represents return for firm $i$ bonds over the time horizon $t$ is $\hat r_i=c_i-(1-R_i)(1-\hat n_i(t))$. In this model the return $\hat r_i$ only takes on two possible values and hence is extremely far from normally distributed. Other sources of risk which we have not included\footnote{For example, changes in firm $i$'s credit quality which do not result in default.} smooth out the probability distribution for $\hat r_i$}. If the assets in this portfolio are uniformally distributed amongst $n$ bonds then  Eq.~(\ref{skewvan}) implies that the portfolio return's skewness goes to zero as $n \rightarrow \infty$. To get a non-zero value for the portfolio return skewness in this limit one must include in the joint survival probabilities terms second order in the asset correlations\footnote{For a discussion of the importance of skewness for portfolio allocation to corporate bonds see [14].}.

The default correlation $d_{ik}(t)$ between firms $i$ and $k$ is defined by 
\begin{equation}
d_{ik}(t)={P_{i,k}(t)-P_{i}(t)P_{k}(t) \over \sqrt{(1-P_i(t))P_i(t)} \sqrt{(1-P_k(t))P_k(t)}}.
\end{equation}
At linear order in the asset correlations the joint survival probabilities can be expressed in terms of default correlations,
\begin{equation}
P_{1,\ldots, n}(t)=P_1(t) \cdots P_n(t)\left[1+{1 \over 2} \sum_{i \ne j} d_{ij}(t){\sqrt{(1-P_i(t)) (1-P_j(t))}\over  \sqrt{ P_i(t) P_j(t)}}\right].
\end{equation}

In the case where all the off diagonal elements of the correlation matrix are taken to be the same $\xi_{ij}=\xi$, for $i \ne j$ we can define the joint survival probability correlation duration,
\be
D_{1, \ldots, n}(t)={1 \over P_{1, \ldots, n}(t)} {{\rm d} P_{1, \ldots, n}(t) \over {\rm d} \xi},
\ee
which characterizes the sensitivity of the joint survival probability to changes in the asset correlations. Our results give an explicit formula for the joint survival probability correlation duration evaluated at zero correlation,
\be
\label{duration}
D_{1, \ldots, n}(t)|_{\xi=0}={1 \over 2}\sum_{i \ne j}{1 \over P_{i}(t) P_{j}(t)} \int_0^t{\rm d}t' \sigma_i\sigma_j A_{ij}(t,t').
\ee

Consider five hypothetical firms with identical properties $\sigma_i=0.30$ (annualized), $d_i/V_{i0}$=0.30 and  $\xi_{ij}=\xi$ , for $i \ne j$. Here we calculate their risk free joint survival probability and so we set $\mu_i=r$. Finally we take $\lambda_i=r$ so their default barriers grow at the risk free rate (we also assume no dividend payments, $q_i=0$). Then we find that $P_i(5 \rm{yr})=87.3\%$ and so $P^{(0)}_{1,2,3,4,5}(5 {\rm yr})=P_i(5 \rm{yr})^5=50.6\%$. The effects of correlations are calculated by performing the integrals in Eq.~(\ref{mainresult1}) and (\ref{mainresult2}) numerically. The joint survival probability correlation duration is $D_{1,2,3,4,5}(5{\rm yr})=10\times 0.30^2 \times 0.611=0.55$. This implies that $P^{(1)}_{1, \ldots, 5}(5 {\rm yr})/P^{(0)}_{1,\ldots, 5}(5{\rm yr})= 0.55\times \xi$. So for an asset correlation $\xi=0.30$ the correlations increase the five year survival probability from $50.6\%$ to $50.6\times(1+0.55 \times 0.30)\%=58.9\%$.
Default correlations increase with the time horizon $t$. For example, Lucas [15] estimates that over one year, two year and five year time horizons default correlations between $Ba$ rated firms are $2\%$, $6\%$ and $15\%$ respectively. Using Eqs.~(\ref{mainresult1}), (\ref{mainresult2}) and (\ref{mainresult3}) we find that with $\xi=0.3$ the five year risk neutral default correlation for any pair of the hypothetical firms considered in this example is $d_{ij}(5{\rm yr})=11.3\%$. We will provide evidence in Section 4 that $\xi=0.3$ is small enough that first order perturbation theory is an accurate approximation. Hence our results are useful for pricing first to default baskets of five firms with realistic values for the correlations.

\section{Asset Correlation Dependence of The Joint Survival Probability For Five Industrials}

As a practical example of the results of the previous section we calculate here the dependence of the risk neutral joint survival probability of five basic industrials on their asset correlations. The five companies we consider are: Alcoa Inc. (ticker AA), Dow Chemical Company (ticker DOW), E.I. du Pont de Nemours and Company (ticker DD), International Paper Company (ticker IP) and Weyerhauser Company (ticker WY). 

For these companies we take $d_i=D_{i,0}/V_{i0}$ where $D_{i0}$ is the present value of the total debt. This corresponds to a recovery fraction on the total debt very near unity. Note that the recovery fraction on the total debt can be much greater than on a particular bond issue since the recovery fraction on a particular bond issue depends on its level of subordination. Sometimes practitioners take the recovery fraction to be random [16] reflecting uncertainties associated with accounting transparency and other issues\footnote{It is possible to generalize the results of this paper to the fluctuating default threshold case. Then Eq.~(\ref{mainresult2}) would contain two additional integrations over values of $d_i$ and $d_j$ weighted with the probability distributions for these companies default thresholds.}. The initial asset value is $V_{i0}=S_{i0}+D_{i0}$ where $S_{i0}$ is the present market cap. Finally the corporate (annualized) asset value volatilities are determined from the historical stock volatility $\sigma_{i}^{(S)}$(taken over a two year period) via, $\sigma_i= (S_{i0}/V_{i0})\times \sigma_i^{(S)}$. Table I gives the input parameters for each of these companies that are needed to compute the dependence of their joint survival probability on their asset correlations. The last column of Table I gives their five year risk neutral default probabilities calculated using Eq.~(\ref{probint}). Note that the dividend yield in the fourth column is with respect to the asset value and not the equity value. We calculate the risk neutral probabilities so the drifts $\mu_i$ are set equal to the risk free rate. Finally we assume the default thresholds grow at the risk free rate (i.e., $\lambda_i=r$) and hence the survival probabilities are independent of the value of the risk free rate.

The small value for the default probability for E.I. du Pont de Nemours and Company is a consequence its very small default threshold. There may be items (e.g. legal liabilities ) that  effectively increase the value of $di/V_{i0}$ and are not reflected in our estimate.
\begin{center}
Table I: Input parameters and five year risk neutral default probabilities for five industrials. 
\vskip0.25in
\begin{tabular}{ccccc}
\hline
${\rm Ticker}$ & $di/V_{i0}$ & $\sigma_i$ & $q_i$ & $1-P_i(5{\rm yr})$  \\ \hline
{\rm AA} & 0.19 &  31.2\% & 1.5\% & 4.7\% \\
{\rm DD} & 0.089 &  25.2\% &2.9\% & 0.02\%  \\
{\rm DOW}& 0.24 &  25.0\% & 2.6\% &  3.6\% \\
{\rm  IP} & 0.39 & 16.5\% & 1.4 \%&  2.6\%   \\
{\rm WY} & 0.47 & 16.5\% & 1.4 \% & 8.3\%  \\
\end{tabular}
\end{center}
\vskip0.25in

Our goal in this section is to calculate the dependence of the risk neutral five year joint survival probability on the asset correlations. Since all the companies are in the same sector we assume their asset correlations are the same $\xi_{ij}=\xi,$ for $i \ne j$. Then the joint survival probability correlation duration $D_{1, \ldots 5}( 5{\rm yr})$, evaluated at zero correlation, characterizes this dependence. Using the parameters in Table I and Eq.s~(\ref{mainresult2}), (\ref{mainresult3}) and (\ref{duration}) we find that for the five companies considered here, 
\be
D_{1, \ldots ,5}(5{\rm yr})|_{\xi=0}=0.036. 
\ee
The five year joint survival probability in the uncorrelated case (i.e., $\xi=0$) is $0.82$ and so the probability of one or more of the companies defaulting in a five year period is
\be
\label{example}
1-P_{1, \ldots, 5}(5{\rm yr}) =0.18-0.029\times \xi.
\ee
Since we have worked to linear order in perturbation theory Eq. (\ref{example}) receives corrections of order $\xi^2$. For $\xi=0.30$ correlations decrease the probability of one or more defaults happening in five years from $18\%$ to $17\%$. The correlations have  small effect because Weyerhauser Company  has a much larger probability of defaulting in this period than the other firms.

\section{Comparison With Multivariate Normal Copula Function}

In the practitioner literature a multivariate normal Copula function is sometimes used to determine the dependence of joint survival probabilities on asset correlations. A multivariate normal Copula functions gives for the joint survival probability of $n$ companies\footnote{The Copula function contains more information than this. If the thresholds are evaluated at times $t_1$, \ldots $t_n$ then it gives the joint survival probability, when company $i$ survives to time $t_i$. Here we focus on the case where all the times are the same $t_i=t$.},
\be
\label{copula}
P_{1, \ldots ,n}(t)={1 \over (2\pi)^{n/2}\sqrt{{\rm det} \xi}} \int_{\chi_1(t)}^\infty {\rm d}x_1 \cdots \int_{\chi_n(t)}^\infty {\rm d}x_n{\rm exp}\left[-{1 \over 2}\sum_{ij}x_i\xi_{ij}^{(-1)}x_j\right]
\ee
where, $\xi_{ij}^{(-1)}$, are elements of the inverse of the $n \times n$ asset correlation matrix. The thresholds are determined from the analogous formula for the survival probability of a single firm $i$.
\be
P_{i}(t)={1 \over \sqrt {2\pi}}\int_{\chi_i(t)}^\infty {\rm d}x_i {\rm e}^{-x_i^2/2}
\ee
which gives
\be
\label{copthresh}
\chi_i(t)=\sqrt{2} {\rm erf}^{-1}(1-2P_{i}(t))
\ee
where ${\rm erf}^{-1}$ denotes the inverse of the error function (i.e., ${\rm erf}^{-1}({\rm erf}(x))=x$)
Expanding in the correlations we get for the case of a multivariate normal Copula function that the correlations cause a first order correction to the survival probability given by the simple expression,
\be
P^{(1)}_{1, \ldots, n}(t)/P^{(0)}_{1, \ldots, n}(t)={1 \over 4 \pi}\sum_{i \ne j}\xi_{ij}{1 \over P_{i}(t) P_{j}(t)}{\rm e}^{-\chi_i(t)^2/2} {\rm e}^{-\chi_j(t)^2/2}.
\ee

To study the difference between the value for survival probability with a multivariate normal Copula function a first passage model we consider a hypothetical situation where all the companies have the same properties (i.e., the volatilities, default thresholds and initial asset values, and hence the probabilities of default). Then we can write, in either the first passage or multivariate Copula function models
\be
P^{(1)}_{1, \ldots, n}(t)/P^{(0)}_{1, \ldots, n}(t)={ n(n-1) \over 2} \xi {\cal A}(t)
\ee
where as before we have set $\xi_{ij}=\xi$ for $i \ne j$ and ${\cal A}$ does not depend on the correlations. 

\vskip0.25in
\begin{center}
Table II: ${\cal A}(5{\rm yr})$ in a first passage model and using a multivariate normal Copula function for various choices of parameters. 
\vskip0.25in
\begin{tabular}{cccccc}
\hline
$\sigma_i$ & $di/V_{i0}$ & $P_i(5{\rm yr})$ & $\chi_i(5{\rm yr})$ & ${\cal A}_{\rm fp}(5{\rm yr})/\sigma_i^2$ & ${\cal A}_{\rm C}(5{\rm yr})/\sigma_i^2$ \\ \hline
0.30 & 0.20 &  96.5\% & -1.81 &  0.0697 & 0.0717\\
0.30 & 0.30 &  87.2\% & -1.14 &  0.611 & 0.636\\
0.30 & 0.40 &  73.8\% & -0.636 &  2.06 & 2.17\\
0.35& 0.20 &  91.6\% & -1.38 &   0.223 & 0.231 \\
0.35 & 0.30 & 78.5\% & -0.789 &  1.08 & 1.13  \\
0.35 & 0.40 & 63.4\% & -0.343 & 2.71 & 2.87 \\
\end{tabular}
\end{center}
\vskip0.25in

We compute ${\cal A}$ for several examples using a multivariate normal Copula function and using a first passage model with correlated random walks and compare them. The results are presented in table II. The thresholds $\chi_i$ in the multivariate normal Copula case are determined using Eq.~(\ref{copthresh}) which sets the probability of survival for each company equal to what it is in the first passage model. In columns five and six of table II we present values of ${\cal A}$ (divided by the volatility squared) using the first passage model (${\cal A}_{\rm fp}$) and the multivariate normal Copula function model (${\cal A}_{\rm C}$). In the first passage case we assume the default threshold grows at the risk free rate and that there are no dividend payments. Since we are computing the risk neutral joint probability of survival the drifts $\mu_i$ are set equal to the risk free rate\footnote{Then the survival probabilities don't depend on the risk free rate and so we don't have to assume a value for it.}. The first two columns give the values of the annualized volatility and initial default threshold divided by asset value we use. Column three gives the resulting five year survival probability for a single company and column six gives the threshold in the multivariate normal Copula function that corresponds to this survival probability (computed using Eq.~(\ref{copthresh})). Note that the values of ${\cal A}$ agree to within about $10\%$ in these two models and for the same correlations the multivariate normal Copula function gives a larger joint survival probability than the first passage model. The agreement between the multivariate normal Copula function and the first passage model is better for higher quality firms that have a larger survival probability.

Since the multivariate Copula function allows us to compute the survival probability for any correlation we can use it to see over what range of asset correlations first order in perturbation theory applies. First consider the parameters used in the first line of table II (i.e., $\chi_i=-1.8102$). In that case we find that perturbation theory gives $P_{1, \dots 5}(5{\rm yr})=0.837+0.054\xi$. On the other hand explicit evaluation of Eq.~(\ref{copula}) gives for $\xi=0.1,0.2,0.3,0.4 $ and $0.5$ the joint survival probabilities $P_{1, \dots 5}(5{\rm yr})= 0.842,0.849,0.858,0.867,$ and $0.877 $ respectively. Since in this case $P_{1, \dots 5}(5{\rm yr})$ is close to unity comparing linear perturbation theory with the value of $ 1-P_{1, \dots 5}(5{\rm yr})$ is the best way to measure of its accuracy. For $\xi=0.3$ linear perturbation theory for the probability of one or more of the five companies defaulting during the five year time horizon is accurate to about $3\%$ while for $\xi=0.5$ linear perturbation theory only accurate at the $8\%$ level.
Next consider the parameters used for the second line of table II (i.e., $\chi_i=-1.1383$). In that case we find that perturbation theory predicts $P_{1, \dots 5}(5{\rm yr})=0.5056+0.289\xi$ while explicit evaluation of Eq.~(\ref{copula}) gives for $\xi=0.1,0.2,0.3,0.4 $ and $0.5$ the joint survival probabilities $P_{1, \dots 5}(5{\rm yr})=0.534,0.563,0.591,0.619$ and $0.648$ respectively. Evidently in this case first order perturbation theory is a reasonable approximation for the joint survival probability even if asset correlations are not small. For $\xi\le 0.5$ it is accurate to better than $1\%$. Linear perturbation theory for the joint survival probability works better for companies of lower credit quality that have higher default probabilities.

\section{Concluding Remarks}

We have studied the impact of correlations on joint survival probabilities using a structural first passage model. We derived, using perturbation theory in the correlations, an explicit formula for the impact of asset correlations on the joint survival probability.  We compared these results with what a multivariate normal Copula function predicts for joint survival probabilities and found (in the cases we considered) that the agreement between the two models was quite good for higher quality firms with a low default probability over the time horizon under consideration. Our results use first order perturbation theory which includes effects linear in the asset correlations. 

At first order in perturbation theory the joint survival probabilities can be expressed in terms of default correlations. It is possible to extend our analysis to second order in perturbation theory. 

We argued that for five firms first order perturbation theory provides a useful approximation to the joint survival probability for realistic values of the asset correlations. When the number of firms is very large our results provide a limit where numerical simulations of correlated random walks, with first passage boundary conditions, can be compared with analytic results.

One advantage of structural models is that companies bond and stock properties are related. Using historical stock volatilities we computed the dependence of the joint survival probability for the five basic industrials (i.e.,  Alcoa Inc.,  Dow Chemical Company, E.I. du Pont de Nemours and Company, International Paper Company and Weyerhauser Company) on their asset correlations. 

The methods we use are applicable when the asset correlations are time dependent. One possible extension of the work presented here is to a situation where asset correlations are typically small but for a short period of time they are large. As long as the time period where they are large is small compared with the total time horizon perturbation theory can be used. Periods where correlations are large could result, for example, from market stress associated with heightened geopolitical risk.

\vskip0.25in
\noindent{\Large{\bf References}}
\vskip0.25in

\vspace{0.4cm} \noindent 
[1]~R. Merton, {\it On The Pricing of
Corporate Debt: The Risk Structure of Interest Rates}, Journal of
Finance 29 (1974) 449-470.

\vspace{0.4cm} \noindent 
[2]~F. Black and J. Cox, {\it
Valuing Corporate Securities: Some Effects of Bond Indenture
Provisions}, Journal of Finance 31 (1976) 351-367.

\vspace{0.4cm} \noindent
[3]~T. Ho and R. Singer, {\it Bond Indenture Provisions and the Risk of Corporate Debt}, Journal of Financial Economics 10 (1982) 375-406.

\vspace{0.4cm} \noindent 
[4]~D. Chance, {\it Default Risk and
the Duration of Zero Coupon Bonds}, Journal of Finance 45 (1990) 265-274.

\vspace{0.4cm} \noindent
[5]~R. Jarrow and S. Turnbull, {\it
Pricing Derivatives on Financial Securities Subject to Credit
Risk}, Journal of Finance 50 (1995) 53-85.

\vspace{0.4cm} \noindent
[6]~F. Longstaff and E. Schwartz, {\it A Simple Approach to Valuing Risky Floating Rate Debt}, Journal of Finance, 50 (1995) 789-819.

\vspace{0.4cm} \noindent
[7]~C. Lo and C. Hui, {\it Valuation of Financial Derivatives with Time Dependent Parameters: Lie-Algebraic Approach}, Quantitative Finance 1 (2001) 73-78.

\vspace{0.4cm} \noindent
[8]~M. Wise, P. Lee and V. Bhansali, {\it Corporate Bond Risk from Stock Dividend Uncertainty}, working paper (2003); to appear in the International Journal of Theoretical and Applied Finance.

\vspace{0.4cm}\noindent 
[9]~F. Modigliani and M. Miller, {\it The Cost of Capital, Corporation Finance and the Theory of Investment}, Amer. Econ. Rev. 48 (1958) 261.

\vspace{0.4cm} \noindent 
[10]~R. Merton, {\it On the Pricing of
Contingent Claims and the Modigliani-Miller Theorem}, Journal of
Financial Economics 5 (1977) 241-249.

\vspace{0.4cm} \noindent 
[11]~S. Redner, {\it A Guide to First Passage Processes}, Cambridge University Press, (2001).

\vspace{0.4cm}\noindent
[12]~C. Zhou, {\it An Analysis of Default Correlations and Multiple Defaults}, The Review of Financial Studies, 14, No. 2 (2001) 555-576.

\vspace{0.4cm}\noindent 
[13]~P. Morse and H. Feshbach, {\it Methods of Theoretical Physics}, Part II, McGraw Hill Book Co. (1953).

\vspace{0.4cm}\noindent
[14]~M. Wise and V. Bhansali,{ \it Portfolio Allocation to  Corporate Bonds with Correlated Defaults}, Journal of Risk,
5,1 (2002) 39-58.

\vspace{0.4cm}
\noindent
[15]~D. Lucas, {\it Default Correlation and Credit Analysis}, Journal of Fixed Income, (1995) March, 76-87.

\vspace{0.4cm} \noindent 
[16]~G. Pan, {\it Equity to Credit Pricing}, Risk, November, (2001) 99-102.

\end{document}